\documentclass[11pt]{article}
\usepackage[margin=0.75in]{geometry}
\usepackage{authblk}
\usepackage{graphicx}
\usepackage{amsmath}
\usepackage{amssymb}
\usepackage{bm}
\usepackage{float}
\usepackage{natbib}
\graphicspath{ {figs/} }
\usepackage{subcaption}
\newcommand{\sidecaption}[1]
{\raisebox{\abovecaptionskip}{\begin{subfigure}[t]{1.50em}
  \caption[singlelinecheck=off]{}
  \label{#1}
\end{subfigure}}\ignorespaces}
\usepackage{multicol}
\title{\textbf{1D Anisotropic Surface Wave Tomography with Bayesian Inference}}
\author[1]{John Keith. Magali} 
\affil[1]{Universit\'e de Lyon, UCBL, CNRS, LGL-TPE, 69622 Villeurbanne, France\\ 

Email: john-keith.magali@univ-lyon1.fr
}

\begin{document}

\maketitle

\begin{abstract}
Classically, anisotropic surface wave tomography is treated as an optimisation problem where it proceeds through a linearised two-step  approach. It involves the construction of 2D group or phase velocity maps for each considered period, followed by the inversion of local dispersion curves inferred from these maps for 1D depth-functions of the elastic parameters. Here, we cast the second step into a fully Bayesian probability framework. Solutions to the inverse problem are thus an ensemble of model parameters (\textit{i.e.} 1D elastic structures) distributed according to a posterior probability density function and their corresponding uncertainty limits. The method is applied to azimuthally-varying synthetic surface wave dispersion curves generated by a 3D-deforming upper mantle. We show that such a procedure captures essential features of the upper mantle structure. The robustness of these features however strongly depend on the wavelength of the wavefield considered and the choice of the model parameterisation. Additional information should therefore be incorporated to regularise the problem such as the imposition of petrological constraints to match the geodynamic predictions.
\end{abstract}

\begin{multicols}{2}

\section{Probabilistic approaches to surface wave tomography: An introduction}

Conventional surface wave tomography is usually implemented using a two-step approach \citep[e.g][]{nakanishi1983measurement,nataf1984anisotropy,trampert1995global,romanowicz2002inversion,ritzwoller2002global}. The first step involves the inversion of the arrival times of each period considered in the measured source-receiver dispersion data to infer 2D group or phase velocity maps at a given period. Using the 2D velocity maps, the second step proceeds by inverting a dispersion curve at a given geographical location to estimate the 1D velocity structure beneath this location. One may then build a smooth 3D velocity model by the juxtaposition of the inferred 1D models followed by interpolating them. 

\medskip

The tomography problem is often solved by applying first-order corrections of the forward function $g$ around a reference model $\textbf{m}_0$. Mathematically. this translates to:
\begin{equation}
    \mathbf{d} = g(\mathbf{m_0}) + \frac{\partial g}{\partial m} \Delta \mathbf{ m}.
\end{equation}
Doing so allows it to be treated as a linearised inverse problem. For instance, one may estimate the 2D phase velocity maps by minimizing an objective function containing a data residual term and more than one regularization terms such as:
\begin{equation}
    S=||\mathbf{Gm} - \mathbf{d}|| + R_1 ||\mathbf{m}|| + R_2||\mathbf{Dm}||
\end{equation}
where $G=\frac{\partial g}{\partial m}$ is now a mathematical forward operator which now refers to the linearised physics between the model parameters (in this case surface wave velocities), and the data (dispersion data), $D$ is a second-order smoothing operator, and $R_1$ and $R_2$ are the damping and smoothing parameters, respectively. The last two parameters will often dictate the trade-off between the data misfit (\textit{i.e.}, how well the model parameters predict the data), the proximity of the estimated model from its reference state, and the degree of smoothing in the inverted model. Moreover, the regularization parameters are chosen ad hoc. Hence, the inverted model may be susceptible to non-data driven constraints and that these constraints shroud essential information provided by the data. Lastly, optimization techniques such as this lack the capability to estimate model uncertainties.

The problems associated with non-uniqueness and quantification of uncertainties, coupled by the ever-growing computational capacity of modern supercomputers led to the advancement of probabilistic approaches to geophysical inverse problems, as first exemplified by \citet{mosegaard1995monte}. Following this pioneering study, a volume of studies that involve probabilistic approaches to geophysical inverse problems have been published in seismology \citep[e.g.][]{lomax2000probabilistic,shapiro2002monte,husen2003probabilistic,bodin2009seismic,debski2010seismic,bodin2016imaging}, and rapidly growing in the field of geodynamics \citep[e.g.][]{baumann2014constraining,baumann2015geodynamic,morishige2020bayesian,ortega2020fast}.

\medskip

Casting the inverse problem in a probabilistic framework allows one to utilise the original non-linear mapping $g$ between the data and the model within its forward procedure. However due to the use of sampling-based methods, one is then forced to solve the forward model numerous times depending on the number of candidate models to be sampled. Fortunately, various techniques are available to address such complications that are not solely based on heuristics. Here, we restrict ourselves with Bayesian inference, that is, a form of statistical inference that formulates our solution as an \textit{a posteriori} probability initially based on the information we have prior to  evaluating the inverse problem. The goal is therefore not to create an ensemble of solutions that follow a certain probability distribution \textit{ex nihilo}, but to update a prior probability based on valuable information provided by the data. 

\medskip

In such schemes, the parameter space has to be explored for best possible model candidates that could match the predictions observed at the surface. Grid-search algorithms however are time consuming and thus uniform sampling may not be performed efficiently. As such, direct-search algorithms have been introduced in geophysical inverse problems that sample candidate models within a subset instead of the entire parameter space. A specific class of ergodic algorithms is called Markov chain Monte Carlo (McMC) methods, which has been initially used to solve problems in physics \citep{metropolis1949monte,metropolis1953equation}, but is now widely used in geophysical inverse problems. In this algorithm, the parameter space is randomly sampled based on our current state of knowledge of a given model candidate. The sequence of searching new model candidates depend on an acceptance probability that is often determined by satisfying a detailed balance condition to ensure stationarity of the desired solution, the most common being the Metropolis-Hastings algorithm \citep{hastings1970monte}.

\medskip

In this manuscript, we cast the second step of the surface wave tomography problem in a Bayesian framework. That is, we assume that we have completely inferred 2D phase velocity maps of the regions considered, and invert for 1D velocity structures that explain the dispersion curves built from these maps. The problem is applied to synthetic data for each geographical location considered whose anisotropic signatures are solely influenced by convective flow in the mantle beneath it. The solution is a marginal posterior distribution of 1D velocity models that best explain the data accompanied by their uncertainty limits. We show that even with strong \textit{a priori} constraints, conventional surface tomography falls short to capture the complete picture related to mantle deformation. Still, some of the notable features are resolved more or less. This work builds upon the hypothesis that adding geodynamical and petrological constraints would allows us to reduce the number of acceptable tomographic models that are consistent with the geodynamical predictions. 

\section{Methods and experiments design}
In this section, we discuss the full implementation of a 1D anisotropic surface wave tomography in a full Bayesian parameter search approach. We highlight in full detail the (1) model parameterisation, (2) the forward problem, (3) the data, and (4) the inverse approach. The method is applied to synthetic surface wave dispersion curves produced by simple setups of an intrinsically anisotropic upper mantle.

\subsection{Model parameterisation of a 1D Earth structure}

\subsubsection{Radial anisotropy component}

Surface waves are sensitive to 13 depth parameters which are just a linear combination of the full elastic tensor $S_{ij}$ \citep{montagner1986simple}. In particular, Love and Rayleigh phase velocities are sensitive to five depth depth parameters which make up an azimuthally-averaged vertically transverse isotropic (VTI) medium. These parameters are also known as the Love parameters, and by convention, are designated as $A$, $C$, $F$, $L$, and $N$. Note that these functions are constrained by the isotropic phase velocities, and are independent of the azimuth of surface wave anisotropy. As a supplementary, the seismic wave velocities propagating either parallel or perpendicular to the symmetry axis can be written as:
\begin{align}
    V_{PH} = \sqrt{\frac{A}{\rho}} \\ 
    V_{PV} = \sqrt{\frac{C}{\rho}} \\ 
    V_{SH} = \sqrt{\frac{N}{\rho}} \\ 
    V_{SV} = \sqrt{\frac{L}{\rho}},
\end{align}
where $\rho$ is the density of the medium. Here, it is important to understand that the vertical symmetry axis need not be the fast axis of anisotropy. Hence, the relative magnitude between $N$ and $L$, and $C$ and $A$ are interchangeable. Most anisotropic tomography studies interpret $L > N$ as vertical flow, since to first-order, the direction of shear is presumed to be vertical and thus aligned with vertically propagating $S-$waves \citep[e.g.][]{montagner1994can}. Likewise, $L < N$ is often interpreted in terms of horizontal flow. When the flow has both horizontal and vertical components, then the resulting anisotropy will be ambiguous. This requires resolving the tilt of anisotropy \citep{montagner1988vectorial,montagner1988vectorial2}.

\medskip

We constrain radial anisotropy by using a more compact form related to the Love parameters. For $P-$waves, the strength of radial anisotropy can be expressed as $\phi = C/A$, whereas for $S-$waves, it is given by $\xi=N/L$. Finally, there exists another anisotropic parameter which relates to the ellipticity $\eta = F/A-2L$. The phenomenology of $\eta$ can be understood through the parameter $F$ which controls the velocity along the direction between the fast and slow velocities.

\subsubsection{Azimuthal anisotropy component}
Assuming a slightly anisotropic medium, azimuthal anisotropy in surface waves can be decomposed into two terms that depend on its azimuth $\theta$. These two terms are usually called the 2$\theta$ and 4$\theta$ components, and are small perturbations around the isotropic phase velocities. The 2$\theta$ and 4$\theta$ components are sensitive to eight depth functions 2$\theta$: $G_s, G_c, B_s, B_c, H_s, H_c$, and 4$\theta$: $C_s$, $C_c$ \citep{montagner1986simple}. 

\medskip

We will only work with azimuthal anisotropy in Rayleigh waves. Rayleigh waves are much more sensitive to the 2$\theta$ terms than the 4$\theta$ terms \citep{maupin20151}. Thus, to first-order, we could eliminate the parameters $C_s$ and $C_c$ from the inversions. Surface waves also poorly resolve the parameters $H_s$ and $H_c$ \citep{bodin2016imaging}. Thus, we could reduce the model dimensionality associated with azimuthal anisotropy down to four parameters $G_s, G_c, B_s$, and $B_c$.

\subsubsection{Pseudo-regularisation}
By using compact notations and first-order approximations, we are able to reduce the possible number of parameters to be inverted for. Four of which: $A$, $L$, $\xi$, $\phi$, and $\eta$ are sensitive to Love and Rayleigh phase velocities, and the remaining four: $G_s, G_c, B_s$, and $B_c$ are sensitive to the azimuthal variations in Rayleigh phase velocities. Working with synthetic data allows us to access the correct values of the parameters defining our Earth model. As we wish to compare conventional tomography with geodynamic tomography, here, we will be placing ourselves in the best case scenario. We impose strong \textit{a priori} constraints to our solution in the tomographic problem. These constraints can be regarded as regularisation parameters which limit the regions in the parameter space to search through. Here, we assume that we have the correct values relating to the $P-$wave structure $A$, $\phi$, $\eta$, $B_s$, and $B_c$; and thus, only invert for $S-$wave-related structures $L$, $\xi$, $G_c$, and $G_s$. The list of parameters to/what not to invert for is summarized in Table~\ref{tab:tab1}.
\begin{table*}[ht]
\centering
\caption{Depth functions constrained by surface waves and their azimuthal variations.}
\begin{tabular}{c c c}
\hline
Parameter & Fixed & Inverted for \\
\hline
$A$ & $\checkmark$ &  \\
$L$ & & $\checkmark$ \\
$\xi$ & & $\checkmark$ \\
$\phi$ & $\checkmark$&  \\
$\eta$ &$\checkmark$  & \\
$G_c$ & & $\checkmark$ \\
$G_s$ & & $\checkmark$ \\
$B_c$ & $\checkmark$&  \\
$B_s$ & $\checkmark$&  \\
\end{tabular}
\label{tab:tab1}
\end{table*}

\begin{figure*}[ht]
\centering
\includegraphics[width=0.6\textwidth]{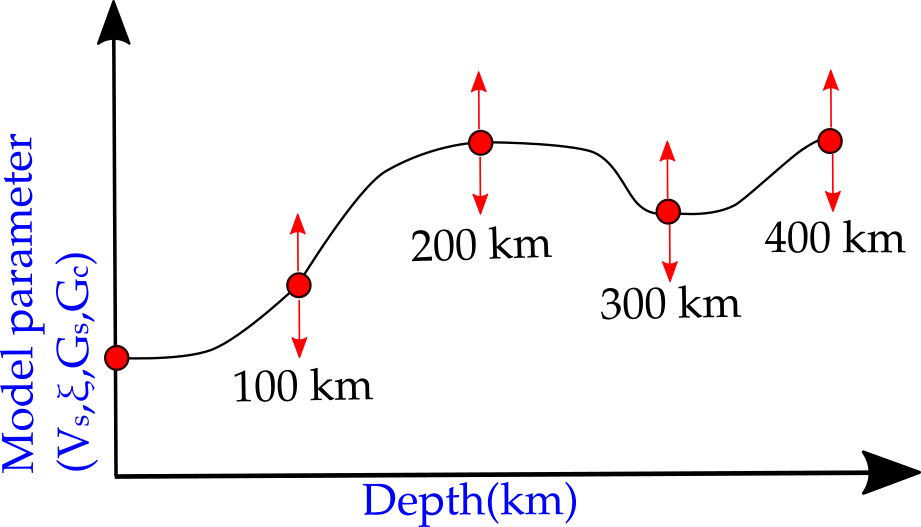}

\caption{Schematic diagram of the 1D parameterisation. The entire region is parameterised with a piecewise cubic Hermite polynomials based on a number of control points. Here, the control points are fixed in depth. The model parameters are then varied at these points using an McMC sampling algorithm. The layers in between are interpolated, and anything below the 400 km is isotropic PREM.}
\label{fig:1dearth}
\end{figure*}

The 1D Earth structure spans from the surface down to a depth of 400 km. It is subdivided into 62 layers of equal thicknesses where each layer has a given value of the nine depth functions (see Table~\ref{tab:tab1}). Thus, the total number of unknowns to be inverted for is 62 $\times$ 4 = 248. Bayesian inversion is therefore a suitable method to treat such an under-determined inverse problem. To reduce the model dimensions further, the mantle is parameterised using a piecewise cubic Hermite polynomials at fixed control points. In the inversions, we vary the unknown parameters $L$, $\xi$, $G_c$, and $G_s$ at the control points and in-between these points, the parameters are interpolated. Below the 400 km, we impose isotropic PREM \citep{dziewonski1981preliminary}. The 1D structure is illustrated in Fig.~\ref{fig:1dearth}

The model vector is thus defined as:
\begin{equation}
    \mathbf{m} = [\mathbf{L},\bm{\xi},\mathbf{G_s},\mathbf{G_c}],
\end{equation}
where bold faces indicate that each parameter is also a vector of size 62.

\subsection{The forward problem}
For each step in the McMC algorithm, the forward problem is evaluated using the proposed model (see Fig.~\ref{fig:1dearth}) as input. The predicted data from this model is then compared with the observed synthetics.

Isotropic Rayleigh $c_R(T)$ and Love $c_L(T)$ dispersion curves are computed using normal mode summation in a spherical Earth \citep{smith1973azimuthal}. Here, the computations are carried out in a fully non-linear approach following the method developed by \citet{saito1967excitation,saito1988disper80}. The software package \verb|DISPER80| \citep{saito1988disper80} takes 1D depth profiles of $V_p$, $V_s$, $\rho$, $\xi$, $\phi$, and $\eta$ as inputs to compute for $c_R(T)$ and  $c_L(T)$ and their associated sensitivity kernels using a Runge-Kutta matrix integration scheme. 

\medskip

Following the pioneering work of \citet{montagner1986simple}, the azimuthal variations in surface wave phase velocities $c_1$ and $c_2$ can be evaluated using the following expressions:
\begin{align}
    c_1(T) = \int_{z=0}^{\infty} \left(   B_c(z)\frac{\partial c_r(T)}{\partial A} +  G_c(z)\frac{\partial c_r(T)}{\partial L}        \right) dz \label{eq:c1} \\
    c_2(T) = \int_{z=0}^{\infty} \left(   B_s(z)\frac{\partial c_r(T)}{\partial A} +  G_s(z)\frac{\partial c_r(T)}{\partial L}        \right) dz. \label{eq:c2}
\end{align}
Eqs.~\eqref{eq:c1} and~\eqref{eq:c2} imply that the azimuthal variations in Rayleigh waves are linearized around the reference VTI model after averaging azimuthally. Such approximations are valid assuming the medium is quasi-anisotropic \citep{montagner1986simple,maupin20151}. 

\subsection{The data}
For each geographical location, we can express the local dispersion curve as the sum of the isotropic dispersion curves and their azimuthal variations giving:
\begin{equation}\label{eq:data}
    c(T,\theta) = c_0(T) + c_1(T)\cos(2\theta) + c_2(T)\sin(2\theta),
\end{equation}
where $T$ is the period, and $\theta$ is the azimuth of the propagating surface wave.

\medskip

For Rayleigh waves, we invert $c_0$, $c_1$, and $c_2$ whereas for Love waves, we only invert $c_0$. For simplicity, we neglect the higher-order terms associated with the elastic parameter $N$. such assumptions are valid due to sparse sampling, low sensitivity, or high noise levels.

\subsection{Quantification of anisotropy}
Anisotropic surface wave tomography is capable of constraining azimuthal and radial anisotropy. The level of radial anisotropy can be quantified through the parameter $\xi$. Conversely, there are a variety of ways to quantify the strength of azimuthal anisotropy, and its fast azimuth $\Psi$. Here, we quantify it in terms of the peak-to-peak anisotropy:
\begin{equation}
    \mathrm{azi} = \frac{2G}{L},
\end{equation}
where $G$ = $\sqrt{G_c^2+G_s^2}$. The azimuth of fast propagation is given by:
\begin{equation}
    \Psi = 0.5\arctan\left(\frac{G_s}{G_c}\right).
\end{equation}

\subsection{1D surface wave tomography with Bayesian inference}
The inverse problem is cast in a full Bayesian procedure where the solution is an ensemble of models distributed according to the posterior pdf $p$(\textbf{m}|\textbf{d$_{obs}$}), accompanied by their uncertainty bounds. In this framework, Bayes' theorem holds:
\begin{equation}
    p(\mathbf{m}|\mathbf{d_{obs}}) \propto p(\mathbf{m})p(\mathbf{d_{obs}}|\mathbf{m}).
\end{equation}
The parameter space is searched using a Markov chain Monte Carlo (McMC) algorithm. To produce reasonable acceptance rates, we employed the adaptive perturbation scheme.

\subsubsection{The likelihood function}
The likelihood function $p$(\textbf{d$_{obs}$}|\textbf{m}) quantifies how well the model parameters fit the observed data. In the context of our problem, it is loosely based on the $L^2$-norm cost function in that it measures the level of misfit between the predictions and the observations. Here, it is essential to distinguish the data residuals related from errors in measurement $\bm{\epsilon_d}$, and from the modeling error due to the use of an incorrect forward model $\bm{\epsilon_g}$. Assuming that the errors are independent and describe a random process, the forward problem can be written as:
\begin{equation}
    \mathbf{d_{obs}} = g(\mathbf{m}) + \bm{\epsilon_d} + \bm{\epsilon_g}.
\end{equation}
Thus, $\mathbf{d_{obs}}$ can also be seen as a random process, and the likelihood distribution can be formulated in terms of the pdf of the data errors:
\begin{equation}
    p(\mathbf{d_{obs}|{m}}) = p(\bm{\epsilon_d} + \bm{\epsilon_g}).
\end{equation}
If we then assume that the errors are uncorrelated and follow a univariate Gaussian distribution with zero mean, and variance $\sigma^2$ where $\sigma^2 = \sigma_d^2 + \sigma_g^2$, we can write the likelihood function as an exponential giving:
\begin{equation}\label{eq:like}
    p(\mathbf{d_{obs}|{m}}) = \frac{1}{(2 \pi \sigma^2)^{N/2}} \exp \bigg[ \frac{-||\mathbf{d_{obs}} - g(\mathbf{m})||^2}{2\sigma^2} \bigg].
\end{equation}
where $N$ is the size of the data vector. Since the goal of any optimization problem is to minimise the $L^2$ cost function, minimising it tantamounts to maximising the probability of the Gaussian likelihood function given by eq.~\eqref{eq:like}.

\medskip

The above formulation can be utilised to construct the likelihood function for surface wave data. For instance, the likelihood function corresponding to a single dispersion measurement can be written as:
\begin{equation}
    p(\mathbf{c_{obs}|m}) =  \frac{1}{(2 \pi \sigma_{c}^2)^{N/2}} \exp \bigg[ \frac{-||\mathbf{c_{obs}} - c||^2}{2\sigma_{c}^2} \bigg],
\end{equation}
where $\mathbf{m}$ is the 1D velocity model described in Fig.~\ref{fig:1dearth}, $N$ is the number of discreet periods, and $\sigma_c^2$ is the estimated variance of the data noise. The likelihood functions of the 2$\theta$ terms can be cast in the same manner.

\subsubsection{The prior distribution}
One of the flexibilities of the Bayesian framework is that it enables one to account for prior information provided that it can be formulated as a probability distribution. The choice of the prior depends on the type of geophysical process we aim to tackle. In the context of seismic tomography, the prior information depends on a range velocity structures that are reasonable for the Earth in general. In practice, the prior information is constrained by existing studies. 

\medskip

At this point forward, we assume minimal prior knowledge and hence, make use of uniform prior distributions with wide bounds. Although we acknowledge that using uniform distributions may be a naive way to setup the prior, working with such a simple distribution would already suffice when demonstrating proofs of concept. Here, we know the exact values of the model parameters, and by imposing wide uniform priors, we are able to assess the efficiency of the method by placing ourselves in the worst case scenario. Indeed, a subject of future work is to consider other forms of the prior distribution, for example non-informative priors or even hiearachical Bayes \citep{malinverno2004expanded}. 

\medskip

Let us now consider a given model parameter $m_i$. The prior $p(m_i)$ is prescribed a constant value over a given range of values defined by $[m_{\mathrm{min}},m_{\mathrm{max}}]$. The prior distribution is thus given by:
\begin{equation}\label{eq:prior}
  p(m_i) = 
 \begin{cases}
  0 & m_i > m_\mathrm{max}, m_i < m_\mathrm{min}\\
  \frac{1}{\Delta m} & m_\mathrm{min}\leq m_i \leq m_\mathrm{max}.
 \end{cases}
\end{equation}
Eq.~\eqref{eq:prior} is interpreted as follows. Suppose that we draw a sample for the specific model parameter $_mi$ from the proposal distribution $q$. If $m_i$ is out of bounds, then the proposal is automatically rejected because the value is not specified by the prior. If $m_i$ is within the prior bounds, then the proposal is accepted with condition based on the the acceptance probability $A$. Choosing narrow bounds therefore imposes hard constraints to the model parameters giving less emphasis to the information provided by the data. 

\medskip

Assuming prior independence, the prior $p(\mathbf{m})$ can be written as a product of 1D priors on each unknown parameter considered giving us:

\begin{equation}\label{eq:pri}
 p(\mathbf{m}) =  \prod_{i=1}^\mathrm{Nlyrs} \bigg[ p(L_i)p(\xi_i)p(Gs_i)p(Gc_i) \bigg],
\end{equation}
where $\mathrm{Nlyrs}$ is the number of layers of the 1D Earth model. Eq.~\eqref{eq:pri} implies that the probability of accepting the transition is automatically zero should one of the parameters be outside their respective bounds.

\subsubsection{The Sampling algorithm}
We use a Markov chain Monte Carlo (McMC) algorithm to search the parameter space for 1D Earth model candidates that could explain the surface wave dispersion measurements. The sampler initiates by randomly drawing a reference model from the prior followed by the evaluation of the likelihood function. Within the Markov chain, the current 1D Earth model is perturbed at their control points (see the red arrows in Fig.~\ref{fig:1dearth} to transition into a new state. This is performed by randomly selecting one of the following set of moves:
\begin{enumerate}
 \item Change the Love parameter $L$ values of all control points according to a Gaussian distribution centered at the current value of $L$.
 \item Change $\xi$ values of all control points according to a Gaussian distribution centered at the current value of $\xi$.
  \item Change $G_s$ values of all control points according to a Gaussian distribution centered at the current value of $G_s$.
   \item Change $G_c$ values of all control points according to a Gaussian distribution centered at the current value of $G_c$.
\end{enumerate}
If the proposed 1D Earth model is within their respective prior bounds, we then solve the forward problem completely. The computed dispersion curves are then compared with the observed synthetics following the evaluation of the likelihood function. The resulting probability is then used to evaluate the acceptance probability via the Metropolis-Hastings algorithm. The outcome of the algorithm determines
whether the proposed model is added to the posterior distribution. Should it be rejected, the current model is counted successively in the next iteration.

\subsubsection{An adaptive perturbation scheme}
\label{subsec:AAPS}
We opted for a more dynamic perturbation, that is, the perturbations vary depending on a given situation. Here, we employ the McMC sampler with an adaptive perturbation scheme based on the acceptance rate. This requires keeping track of the acceptance rate for a given model parameter on the fly.
Let us now denote $N_\mathrm{pop}$ to be the population size, that is, total number of samples in the inversion, and $M$ to be the total number of accepted models within $N_\mathrm{pop}$. The acceptance rate corresponding to the entire population is just:
\begin{equation}
    \mathrm{acceptance} = \frac{M}{N_\mathrm{pop}} \times 100 \%
\end{equation}
Next, we need a sizeable amount of samples $n$ within $N_\mathrm{pop}$ to allow for the relaxation of the acceptance rates, that is, the period at which the acceptance rates are in stable conditions. The population $N_\mathrm{pop}$ are then separated into different cycles (\textit{i.e.}, sample window) that are multiples of $n$. For instance, choosing $n=N_\mathrm{pop}$ just pertains to the acceptance rates of the entire population, whereas choosing $n=1$ would constantly reset the counters for the proposal and the accepted models at every iteration. At the end of every cycle, the current values of the acceptance rates are then used to determine whether to increase or decrease the perturbation. If the acceptance is less than 20$\%$, then the perturbations will be reduced by 25$\%$ of its current value. Likewise, if the acceptance is more than 50$\%$, then the perturbations will be increased by 25$\%$ instead. The counters for the proposal and the acceptance are then reset, and the new acceptance rate is recalculated in the subsequent cycle. As an example, suppose that we have $N_\mathrm{pop}=50000$ samples, and a subset of $n=5000$, we thus having ten cycles. After the first 5000 iterations $l$ (first cycle), the acceptance rate at $l=5001$ will determine whether to increase or decrease the perturbations by 25$\%$. Resetting of the counters will then be ensued regardless. The appraisal of the acceptance rates are then marked after every $n$ intervals,that is, at $l=10000,15000,20000$ and so on.

\section{Synthetic example: Application to a 3D deforming upper mantle}
We perform 1D anisotropic surface wave tomography at 32 different geographical locations. To efficiently implement the inversions for each location, the algorithm is parallellised in such a way that multiple Markov chains can simultaneously search the parameter space independently from one another. Here, one Markov chain is allocated to one processor. At the end of the inversion procedure, the ensemble of models from each chain are then gathered to construct the posterior probability distribution. Since we consider 20 independent chains for each geographical location, and we have 32 locations in total, we need a hefty 32 $\times$ 20 = 640 cores to implement the parallel scheme.

\subsection{Upper mantle flow induced by a sinking anomaly}

\begin{figure}[H]
\centering
\includegraphics[width=0.45\textwidth]{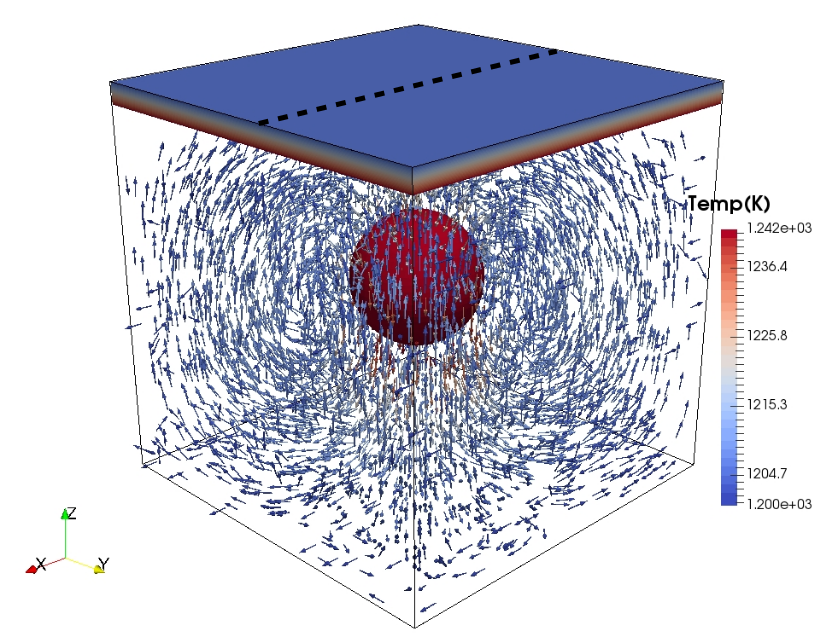}
\caption{Snapshot of a 3D deforming mantle induced by a sinking spherical anomaly. The model domain is of the size 400 km $\times$ 400 km $\times$ 400 km. The isovolumetric gradients correspond to the 3D temperature profile of the region, and the superimposed vector field is the flow induced by the spherical body. The dashed lines at the surface represent the 32 geographical locations of the local surface wave dispersion curves. }
\label{fig:setting}
\end{figure}

We first test the method to a deforming viscous upper mantle induced by a negatively buoyant spherical anomaly, akin to a Stoke's sinker. As illustrated in Fig.~\ref{fig:setting}, the surrounding material responds to the sinking anomaly by producing a return flow. The return flow, together with the downward motion of the anomaly, generates local convection cells whose scales are consistent with that of the upper mantle (which we set at $L_s$ = 400 km). We setup the geographical locations of the local surface wave dispersion curves in such a way that they provide a good coverage of the anomaly. Here, we use 32 locations that slice the region evenly into two (refer to the dashed lines). The observed synthetics are generated entirely by a 3D deforming upper mantle. We refer the reader to \citet{10.1093/gji/ggaa577} for further details.

\begin{figure}[H]
  \centering
  \sidecaption{fig:2dra}
  \raisebox{-\height}{\includegraphics[width=0.45\textwidth]{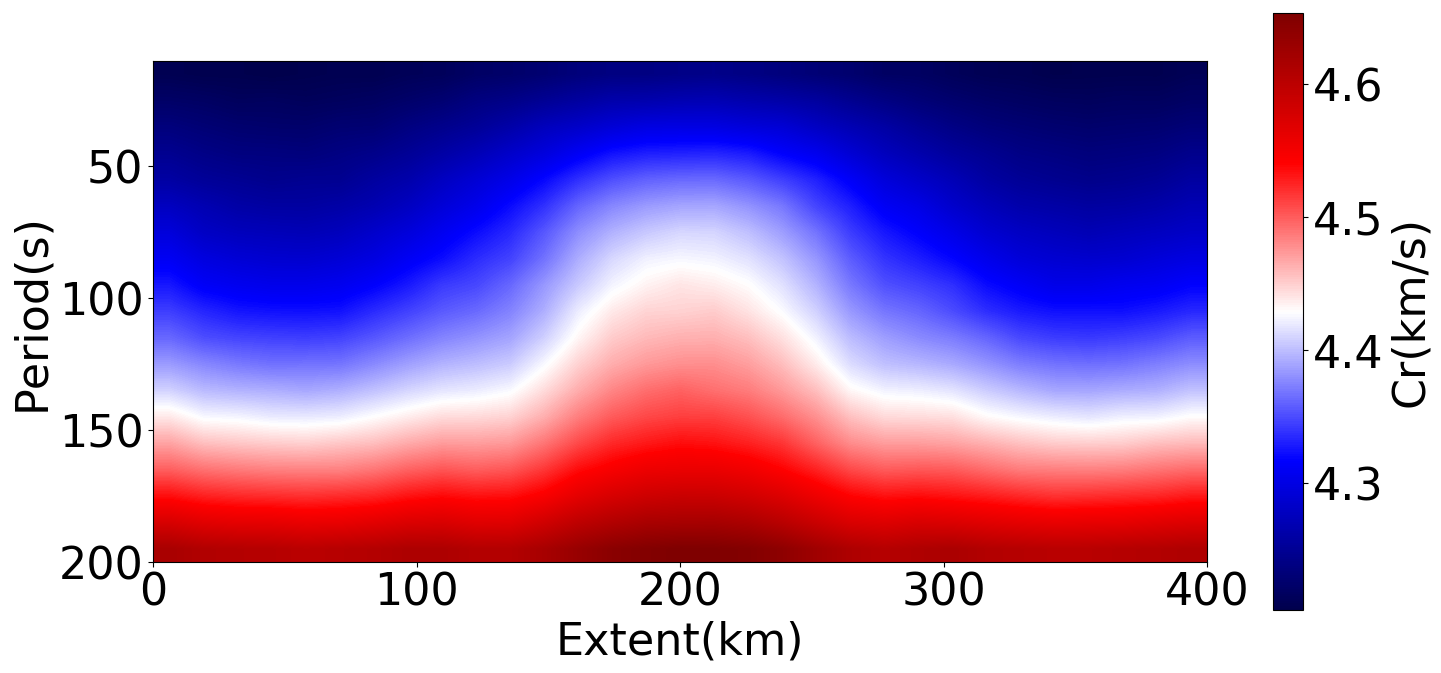}}
  
  \sidecaption{fig:2dlo}
  \raisebox{-\height}{\includegraphics[width=0.45\textwidth]{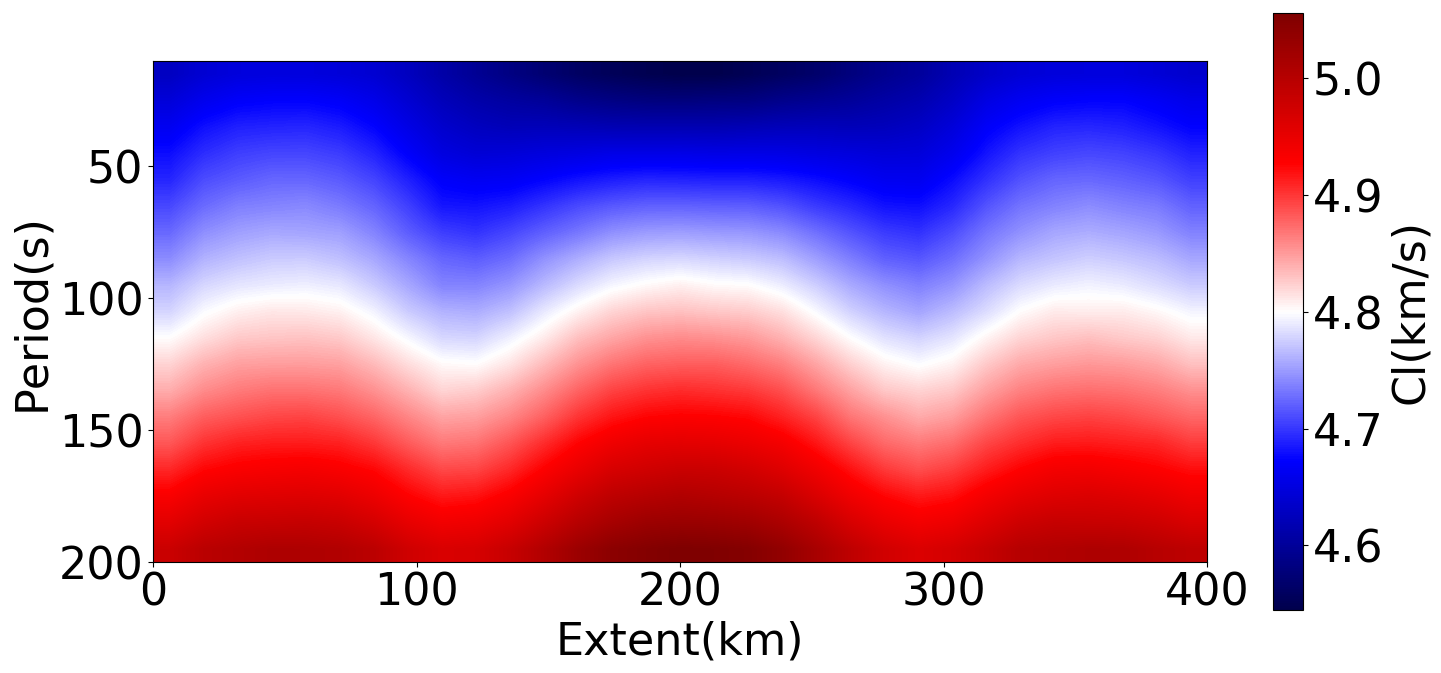}}
  \caption{Reconstructed 2D surface wave maps with added noise. (A) Rayleigh. (B) Love.}\label{fig:2ddisper}
\end{figure}

\begin{figure*}[ht]
  \centering
  \sidecaption{fig:ra}
  \raisebox{-\height}{\includegraphics[width=0.3\textwidth]{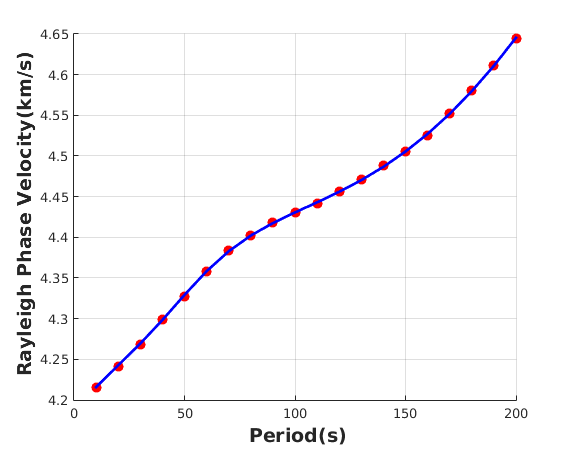}}
  \sidecaption{fig:lo}
  \raisebox{-\height}{\includegraphics[width=0.3\textwidth]{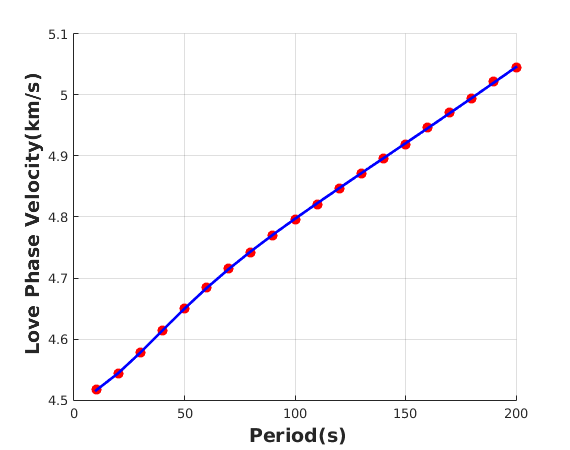}}
  
    \sidecaption{fig:c1}
  \raisebox{-\height}{\includegraphics[width=0.3\textwidth]{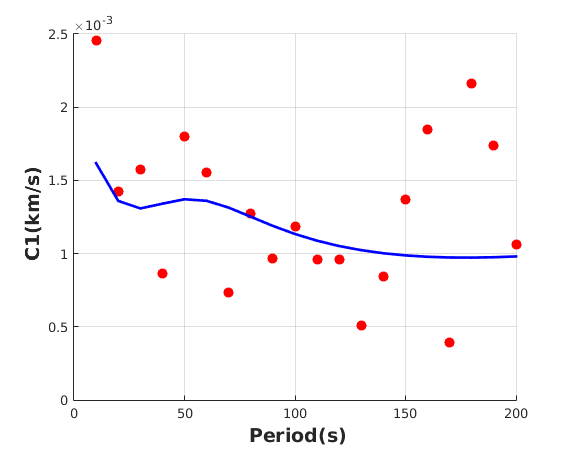}}
  \sidecaption{fig:c2}
  \raisebox{-\height}{\includegraphics[width=0.3\textwidth]{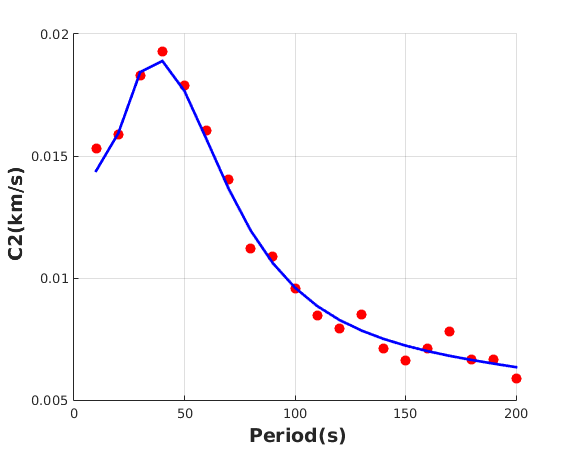}}
 
  \caption{Surface wave phase velocity dispersion curves and their azimuthal variations at a specific geographical location. Solid blue lines are the correct values and the scatter plots are the ones added with noise and are to be inverted.}\label{fig:disper}
\end{figure*}

To mimic real-Earth observations, we tarnish the observed synthetics with noise. We added random uncorrelated noise with standard deviation $\sigma_{c_{R,L}}$ = 0.001 km/s for the isotropic dispersion curves $\mathbf{c_R}$ and $\mathbf{c_L}$, and $\sigma_{c_{1,2}}$ = 0.0005 km/s for the azimuthal components $\mathbf{c_1}$ and $\mathbf{c_2}$. Fig.~\ref{fig:disper} shows the resulting dispersion curves at one given location. Solid blue lines are the synthetic dispersion curves without noise and the red dots represent the synthetics with added random uncorrelated noise. By stacking the dispersion curves laterally (\textit{i.e.}, placing them side-by-side), we are able to visualize the true structure in the data space. Fig.~\ref{fig:2ddisper} shows the effect of the density anomaly onto the resulting surface wave dispersion maps. The negative subsidence observed in the Rayleigh map corresponds to an increases in its speed as it traverses the cold anomaly. 

For each geographical location, the inversion consists of 20 independent Markov chains each containing 1.0 $\times$ $10^6$ samples initiated at a random 1D Earth model drawn from the prior distribution. The prior bounds of each model parameter is summarized in Table~\ref{tab:tab2}.
\begin{table}[H]
\centering
\caption{Prior ranges of the unknown model parameters.}
\begin{tabular}{c c c c c}
\hline
 & $L$(GPa) & $\xi$ & $G_c$(GPa) & $G_s$(GPa) \\
\hline
min & 20 & 0.7 & -5 & -5 \\
max & 150 & 1.3 & 5 & 5\\
\hline
\end{tabular}
\label{tab:tab2}
\end{table}
The models are then collected after a burn-in period of 900,000 samples to ensure the chains are in steady-state and are properly sampling the posterior distribution. Note that even though the models are randomly initiated, the positions of the control points are fixed, and thus may have strong implications on the recovered structures.

\begin{figure}[H]
  \centering
  \sidecaption{fig:ltrue}
  \raisebox{-\height}{\includegraphics[width=0.2\textwidth]{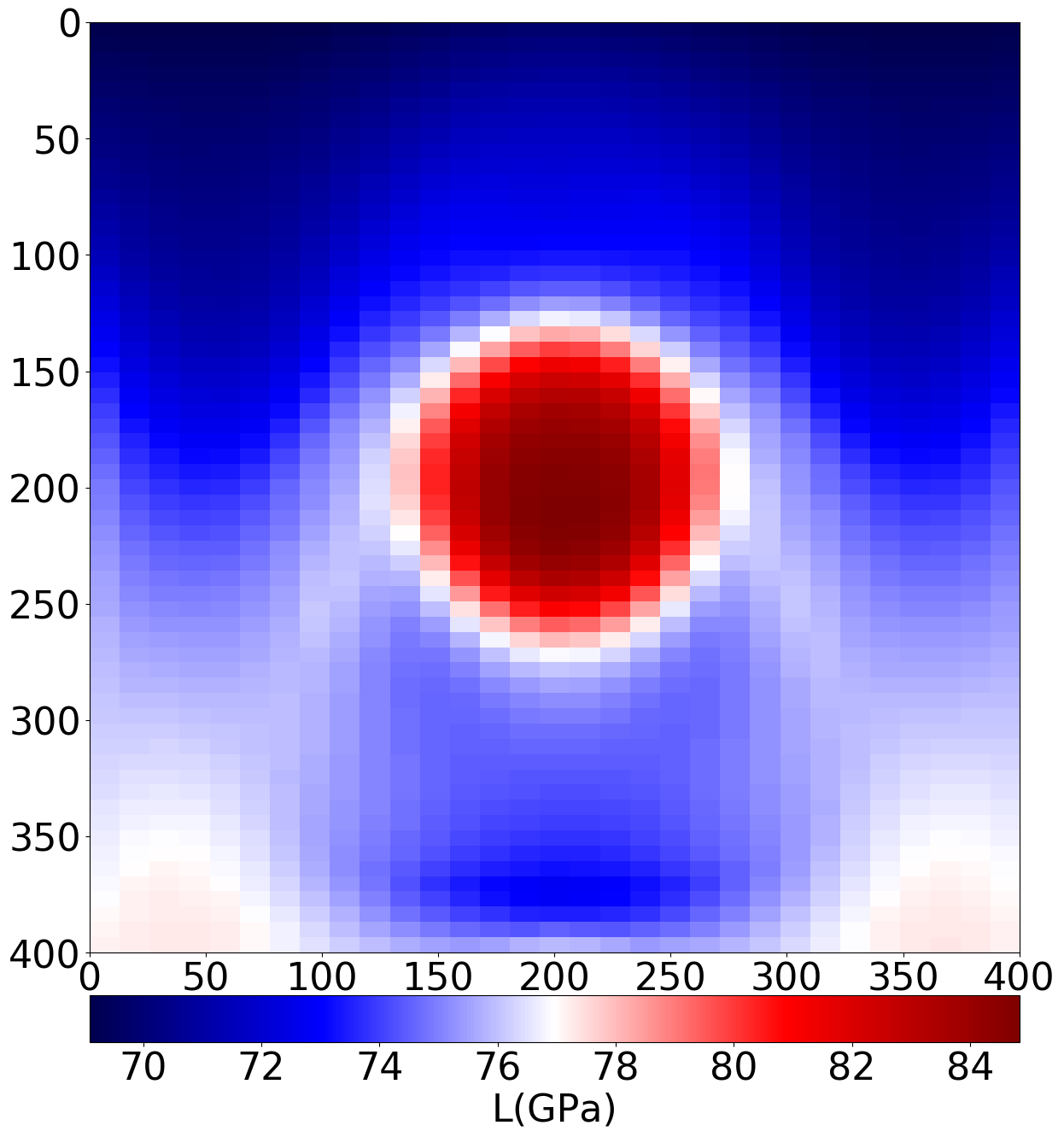}}
  \sidecaption{fig:lmean}
  \raisebox{-\height}{\includegraphics[width=0.2\textwidth]{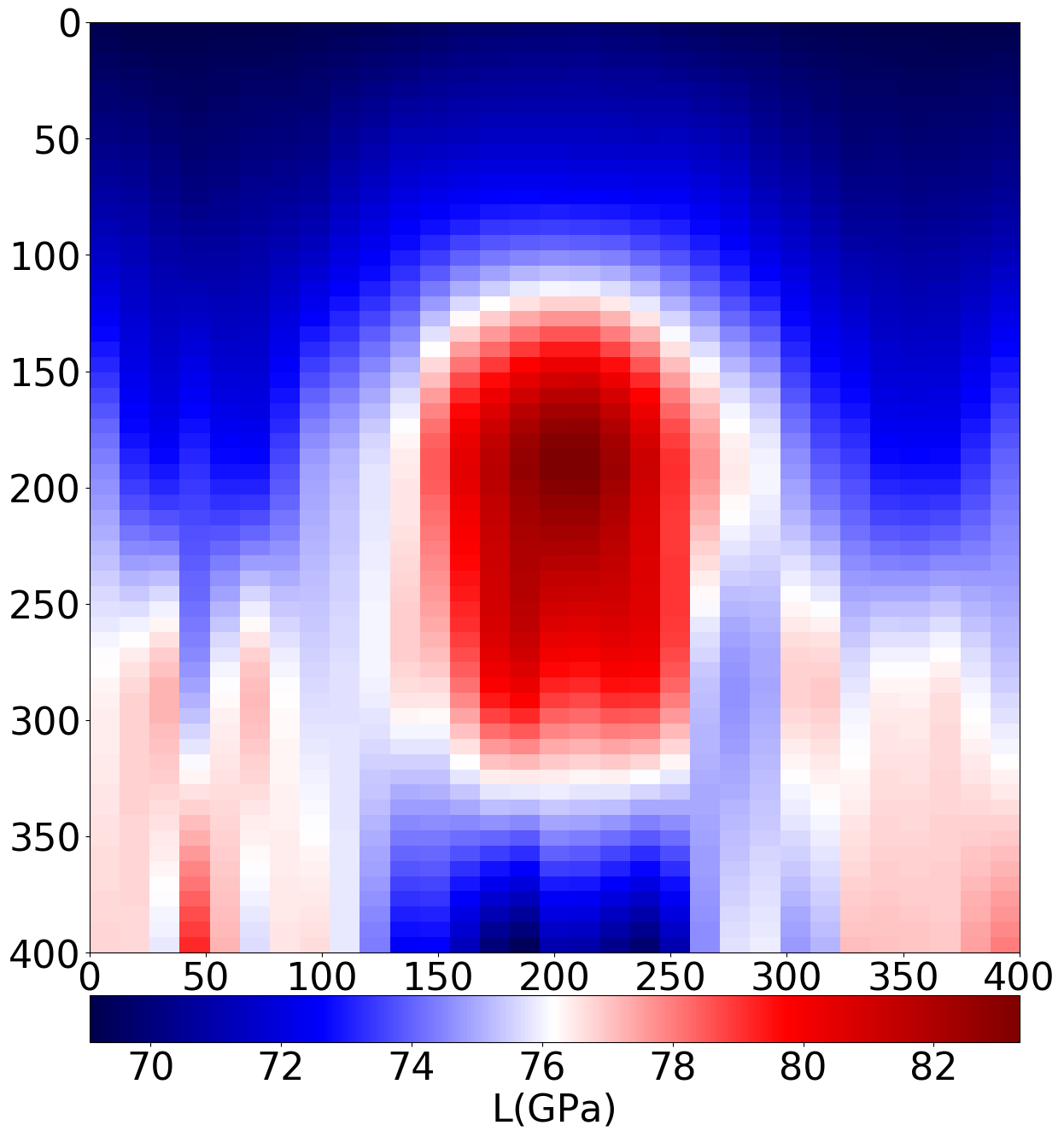}}
  
    \sidecaption{fig:xitrue}
  \raisebox{-\height}{\includegraphics[width=0.2\textwidth]{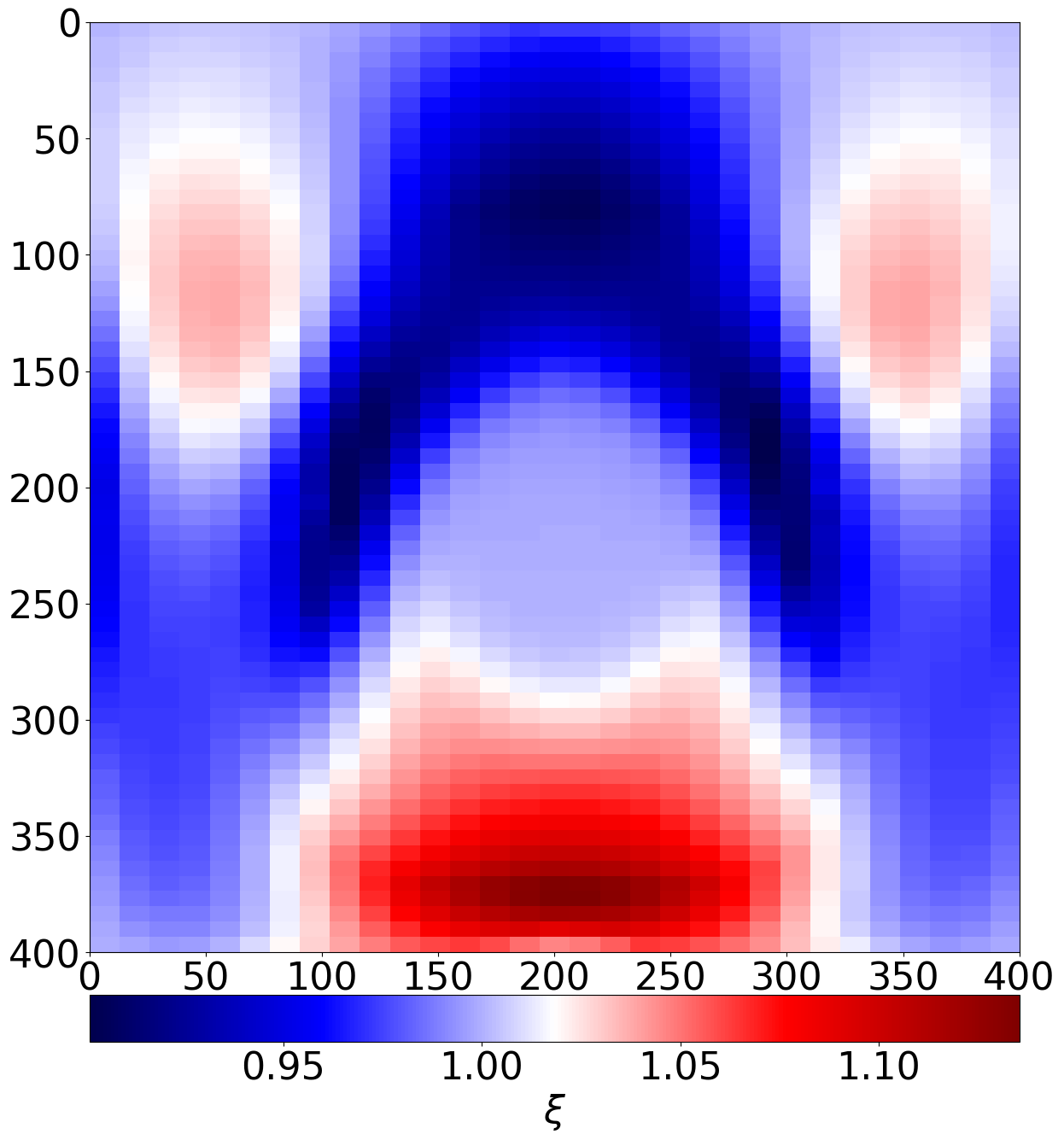}}
  \sidecaption{fig:ximean}
  \raisebox{-\height}{\includegraphics[width=0.2\textwidth]{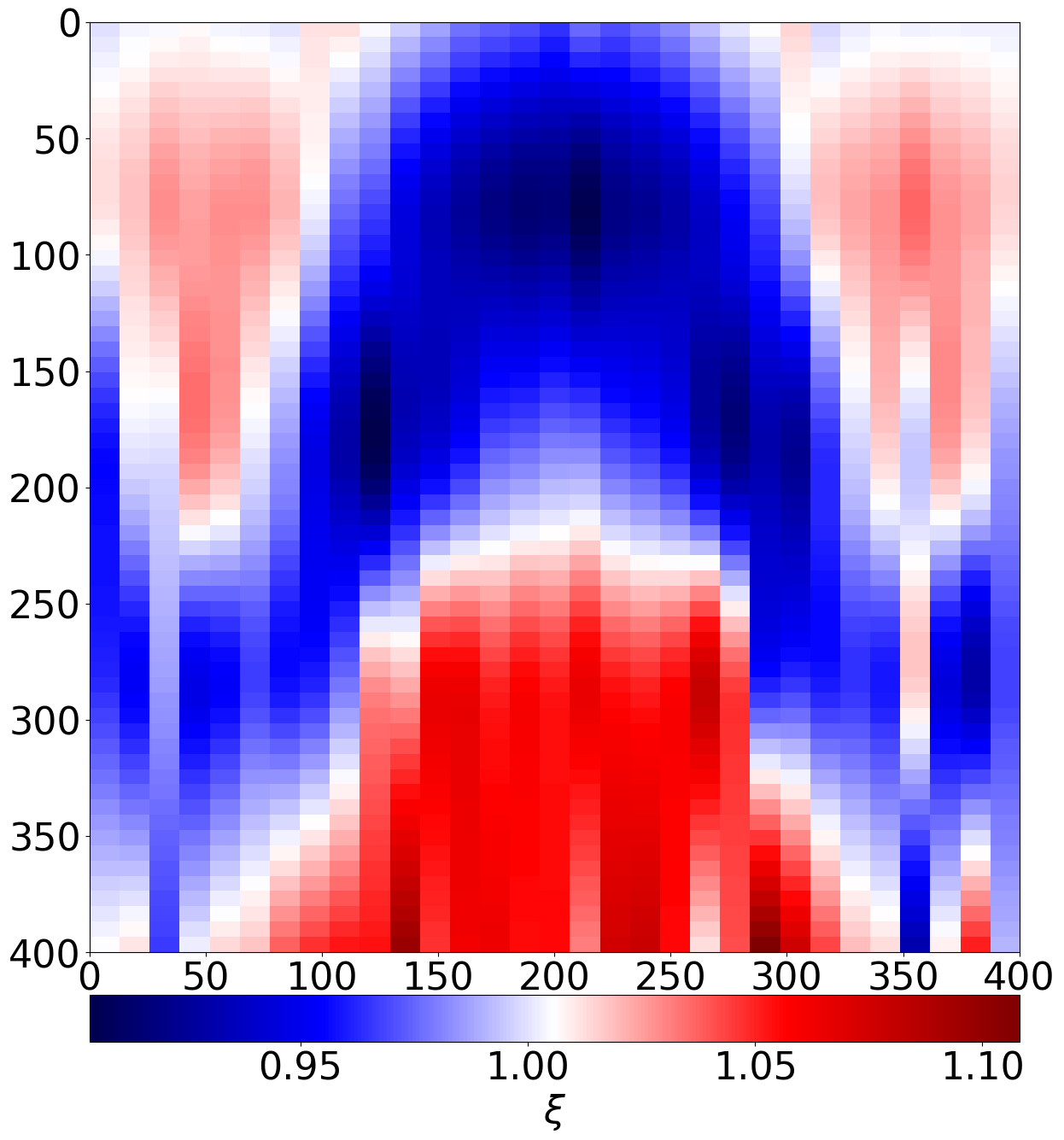}}
 
  \caption{True models (left), Mean models recovered from the inversion (right). (A) and (B) $L-$ structure, (C) and (D) radial anisotropy. }\label{fig:2dmean}
\end{figure}

Fig.~\ref{fig:2dmean} shows the mean velocity structure and the mean radial anisotropy recovered from Bayesian inversion, and the correct structures. The 2D models are constructed by placing the recovered 1D mean structures side-by-side. Surface waves were able to successfully map the most prominent feature, that is, the seismic anomaly associated with the denser sphere. Radial anisotropy was also successfully recovered. However, one of the major drawbacks of surface waves is that they are more sensitive at shallower depths. Hence, the top layers are better resolved than the bottom layers. We thus expect the upper portion to exhibit less model uncertainties that the bottom half. Additionally, some essential features are smeared vertically, which we attribute to the inherent long period nature of surface waves. The structure also appears to be smooth with depth which in part is due to the choice of parameterisation (\textit{i.e.}, cubic splines are smooth functions). The lateral resolution however does not exhibit complete smoothness; instead, appears to be degraded at some regions. This is a result of using randomly uncorrelated data. In this case, the random data noise manifest as small-scale artifacts in the model space. 

\begin{figure}[H]
\centering
\includegraphics[width=0.5\textwidth]{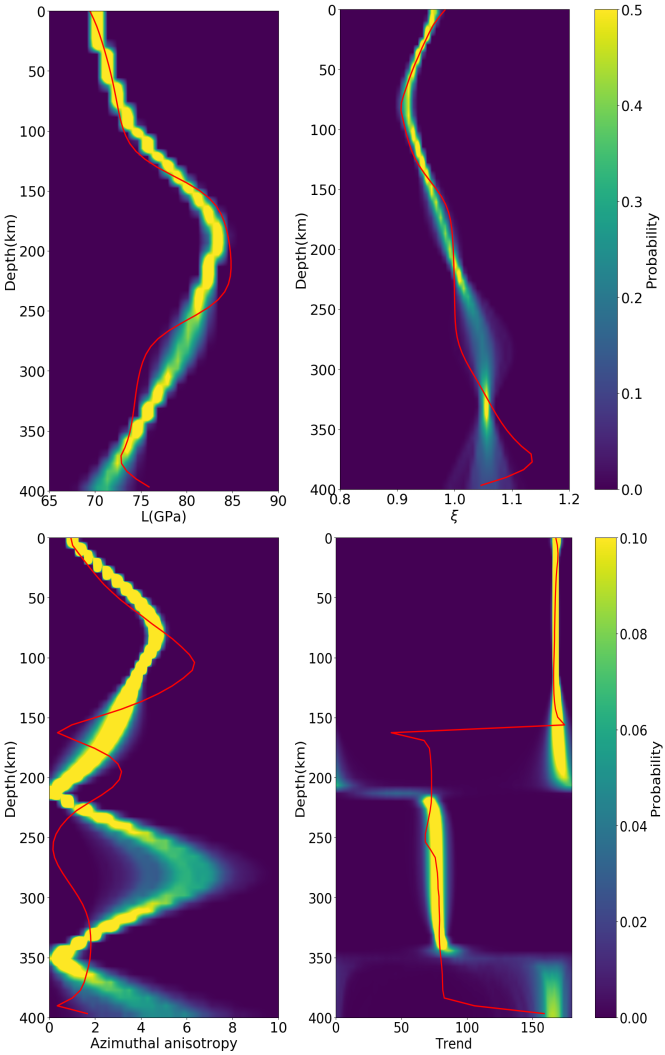}
\caption{1D marginal posterior distributions of $L$, $\xi$, peak-to-peak azimuthal anisotropy, and its fast azimuth $\Psi$ at a specific geographical location, inferred from the Bayesian inversion of surface wave dispersion curves. The true structures are plotted in solid red.}
\label{fig:1dprofile}
\end{figure}

One of the main advantages of a Bayesian framework is we can express the solution as a marginal posterior pdf where the width of the distribution quantifies the model uncertainties. 
These depth profiles were obtained by merging the ensemble of models from the 20 Markov chains. The entire number of models $\mathbf{m}$ used to build the posterior pdf is thus 2.0 $\times$ $10^6$. Results show that the profiles successfully capture the true structure although sharp gradients fail to be resolved. The resulting azimuthal anisotropy shows some peculiarities. Since we only make use of five control points which are fixed, the saddle points of the true azimuthal anisotropy fail to be captured. The increase in model uncertainty at depths is again a result of surface waves being concentrated at shallower depths. Fig.~\ref{fig:1dprofile} shows 1D marginal distributions versus depth of $L$, $\xi$, peak-to-peak azimuthal anisotropy, and its fast azimuth $\Psi$ at a specific geographical location. 

\subsection{Upper mantle flow induced by subduction}
\begin{figure}[H]
\centering
\includegraphics[width=0.5\textwidth]{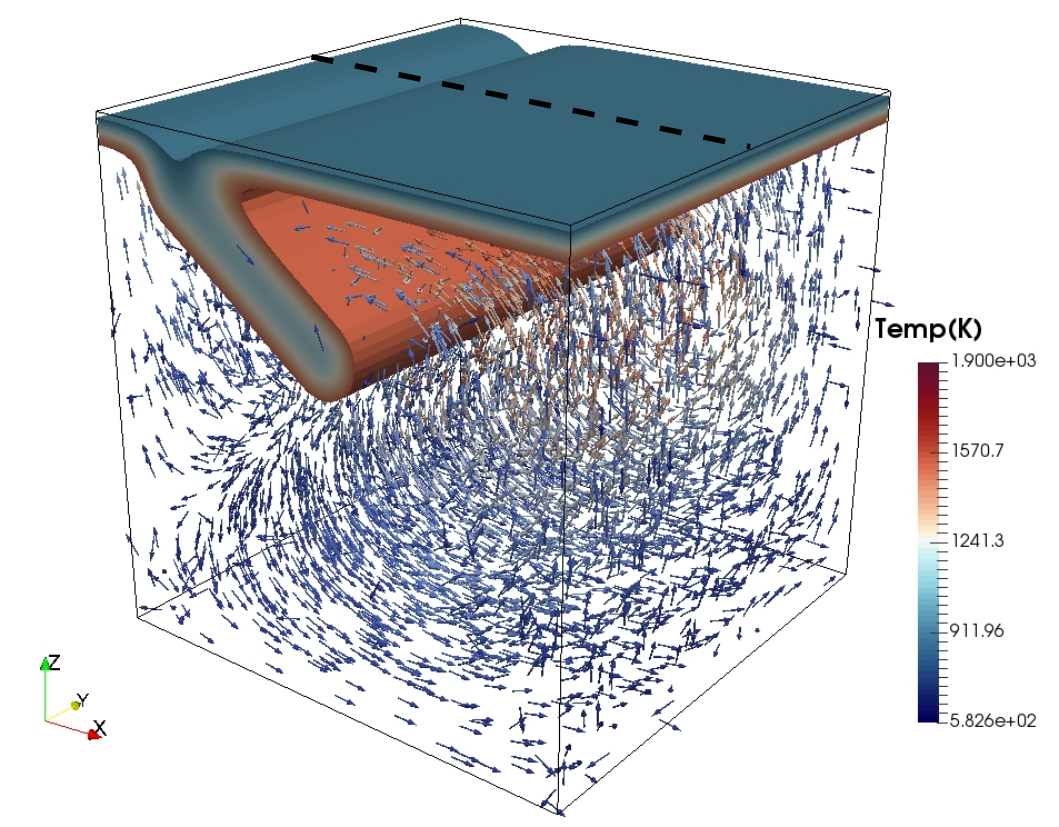}
\caption{Snapshot of a 3D upper mantle flow induced by subduction. The model dimensions are also of the size 400 km $\times$ 400 km $\times$ 400 km. The dashed lines at the surface represent the 32 geographical locations of the local surface wave dispersion measurements. }
\label{fig:setting2}
\end{figure}

\begin{figure}[H]
  \centering
  \sidecaption{fig:2dra2}
  \raisebox{-\height}{\includegraphics[width=0.45\textwidth]{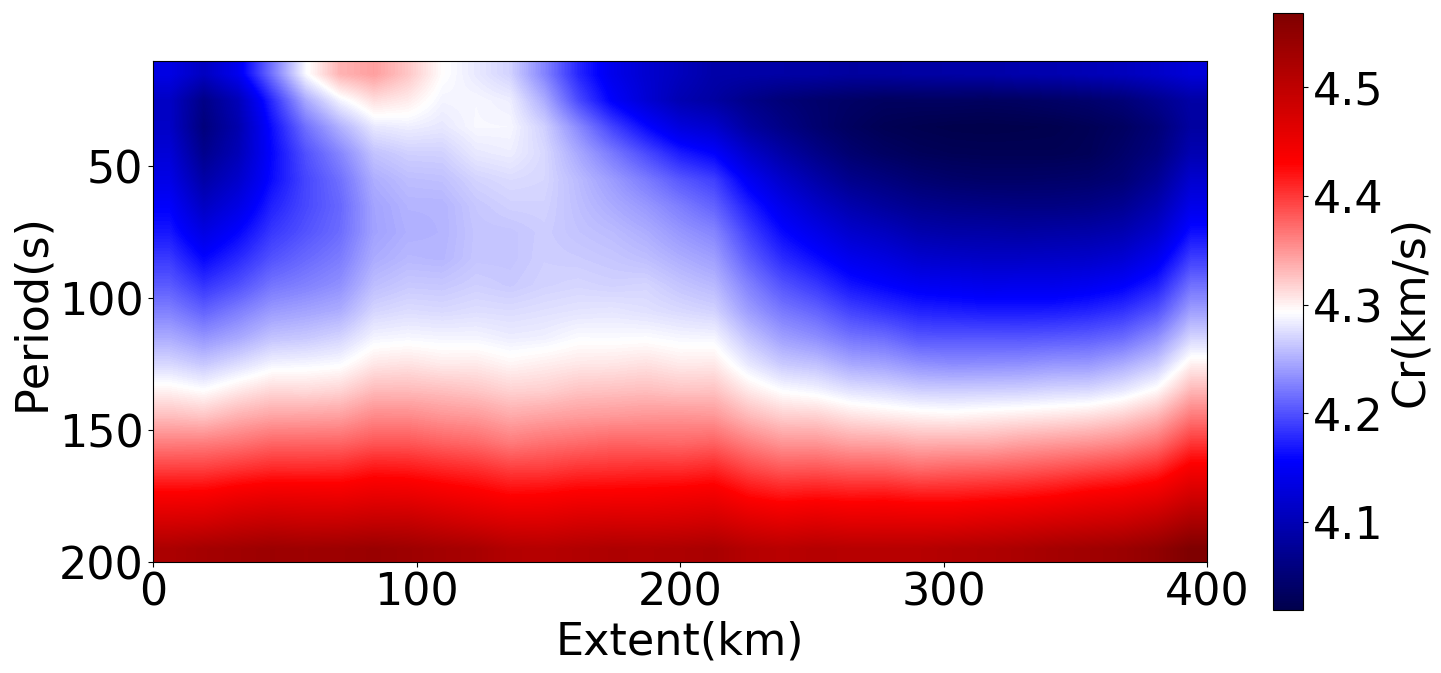}}
  
  \sidecaption{fig:2dlo2}
  \raisebox{-\height}{\includegraphics[width=0.45\textwidth]{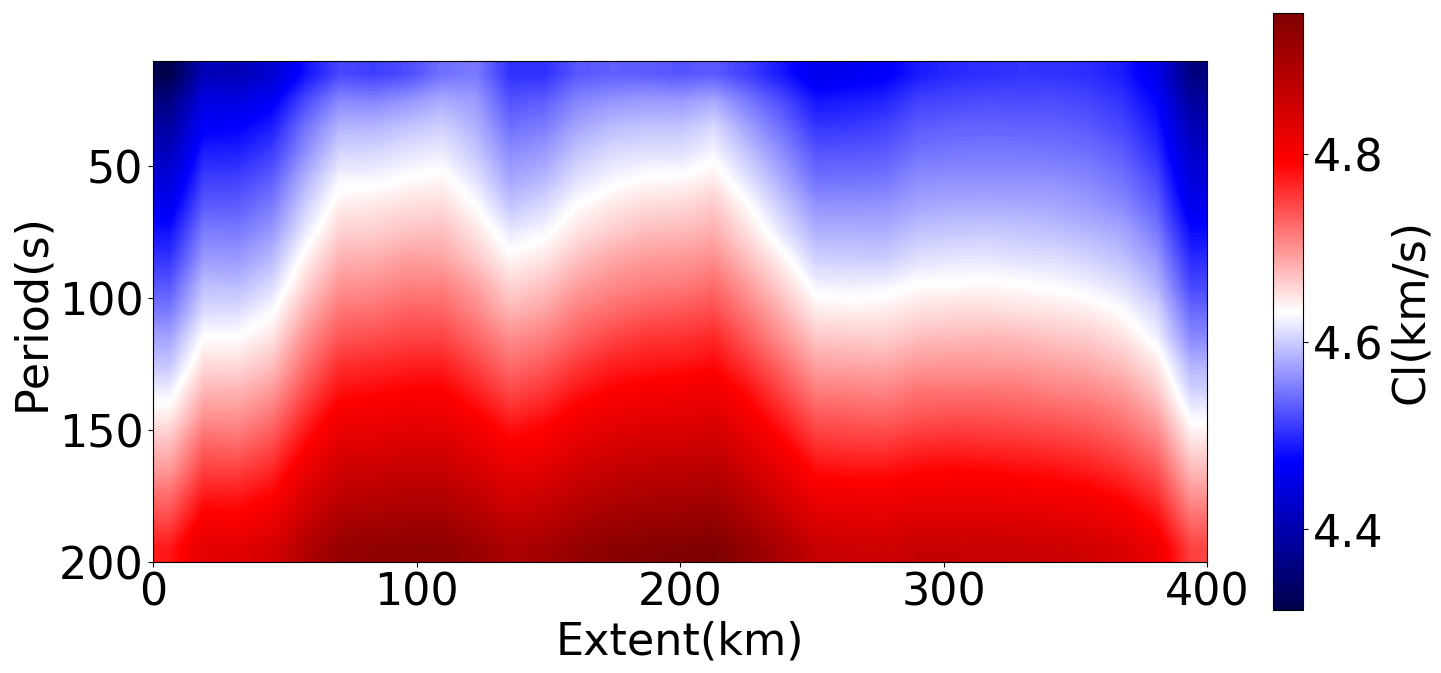}}
  \caption{Reconstructed 2D surface wave maps of a subduction zone with added noise. (A) Rayleigh. (B) Love.}\label{fig:2ddisper2}
\end{figure}

We up the notch further by considering instantaneous flow in the upper mantle induced by subduction as shown in Fig.~\ref{fig:setting2}. Similar to the previous example, we added random uncorrelated noise with $\sigma_{c_{R,L}}$ = 0.001 km/s and $\sigma_{c_{1,2}}$ = 0.0005 km/s. Fig.~\ref{fig:2ddisper2} shows the surface wave maps across a subduction zone. As expected, surface waves (especially Rayleigh in this case) undeniably map the subducting slab.


\begin{figure}[H]
  \centering
  \sidecaption{fig:ltrue2}
  \raisebox{-\height}{\includegraphics[width=0.2\textwidth]{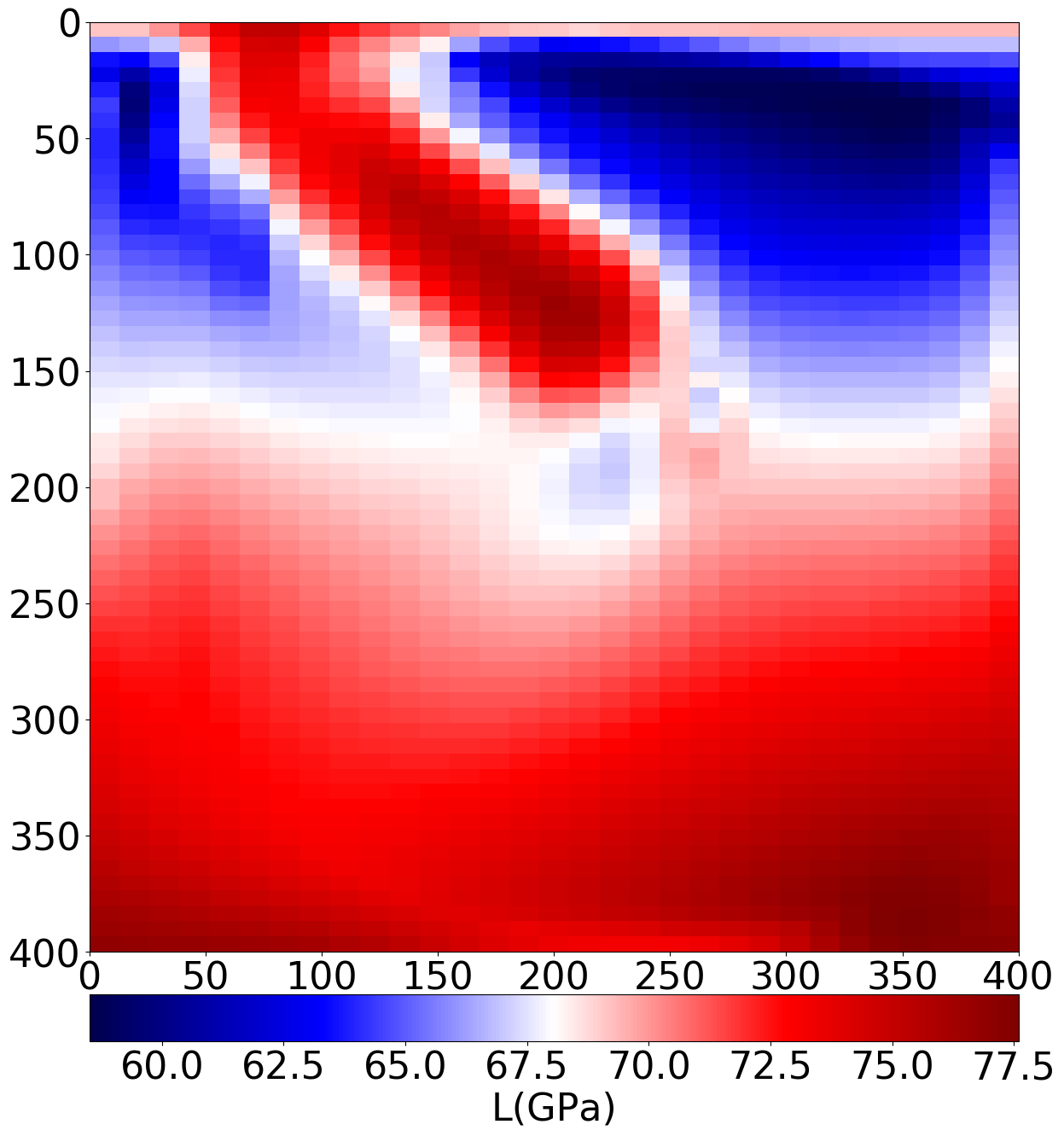}}
  \sidecaption{fig:lmean2}
  \raisebox{-\height}{\includegraphics[width=0.2\textwidth]{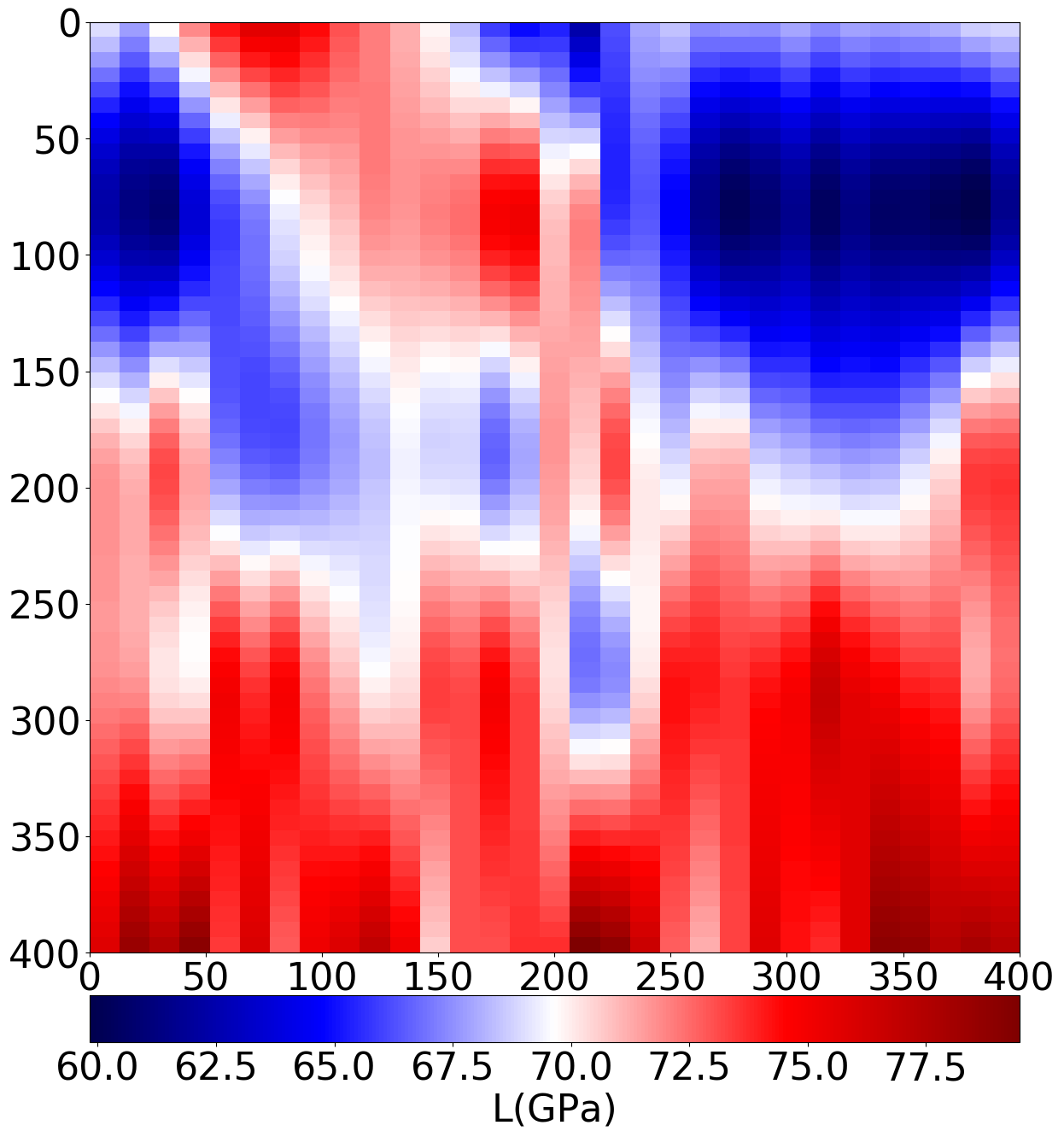}}
  
    \sidecaption{fig:xitrue2}
  \raisebox{-\height}{\includegraphics[width=0.2\textwidth]{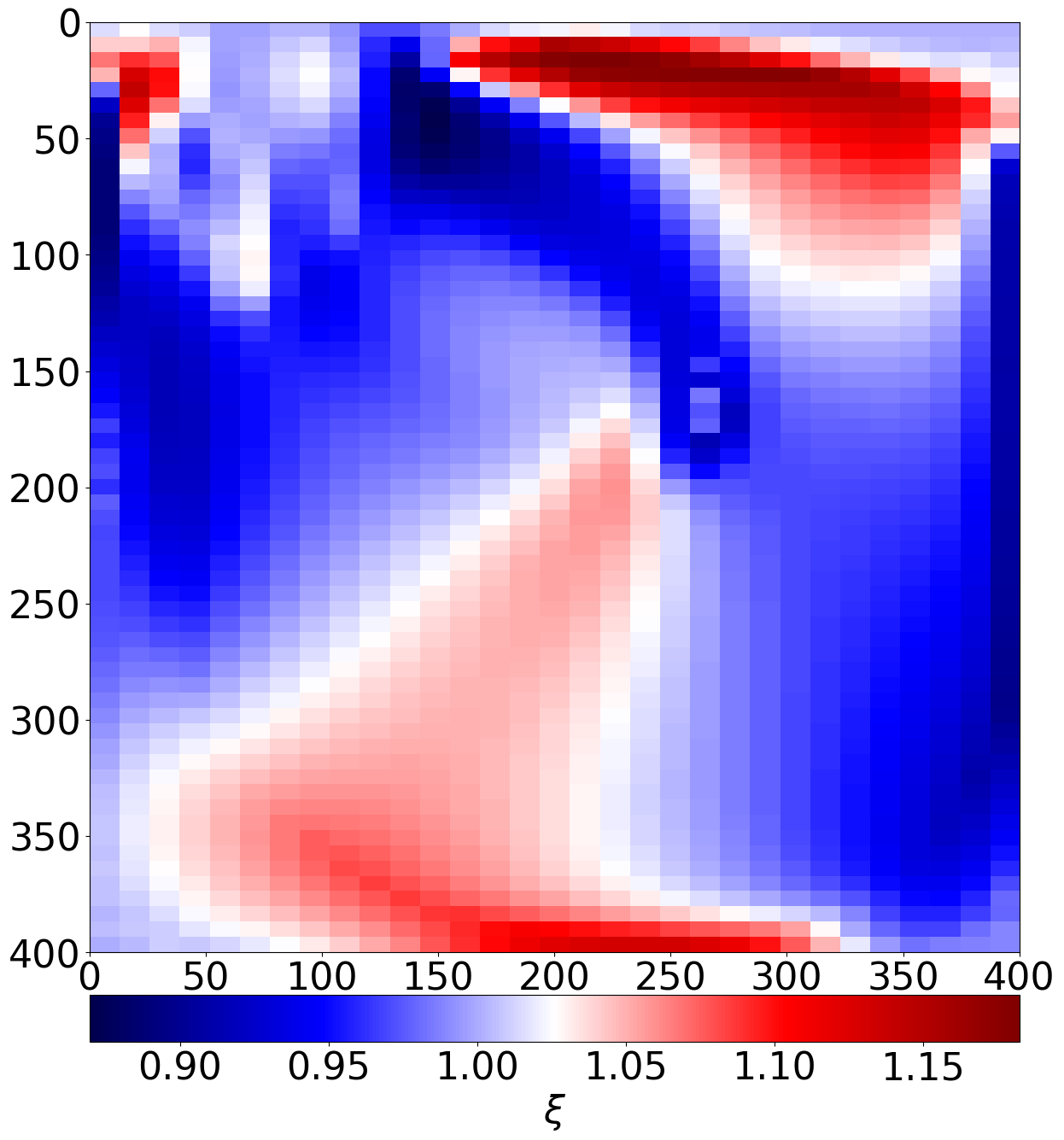}}
  \sidecaption{fig:ximean2}
  \raisebox{-\height}{\includegraphics[width=0.2\textwidth]{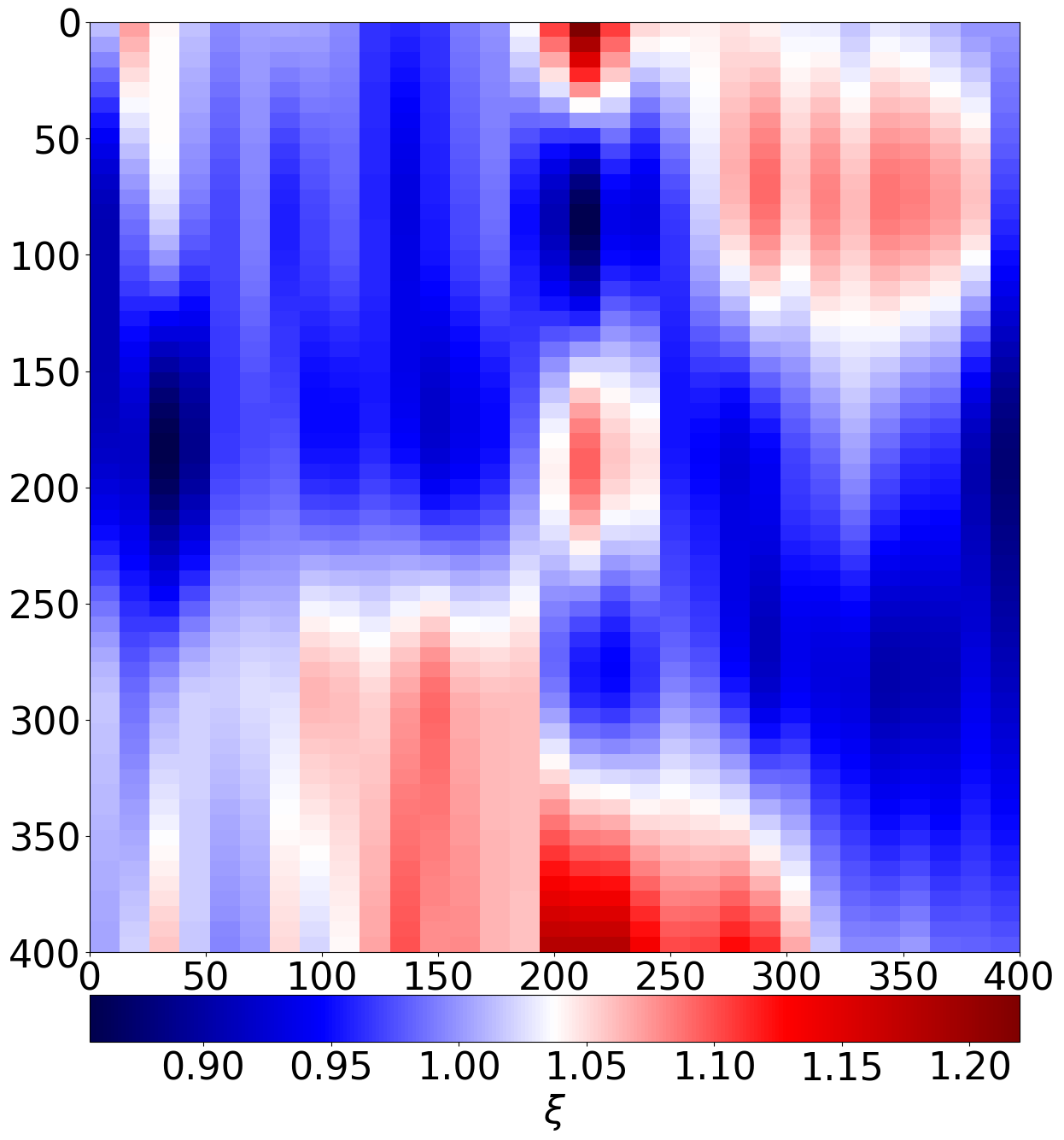}}
 
  \caption{\textbf{Subduction case}. True models (left), Mean models recovered from the inversion (right). (A) and (B) $L-$ structure, (C) and (D) radial anisotropy. }\label{fig:2dmean2}
\end{figure}

The inversion properties are preserved (\textit{e.g.}, number of independent chains, same prior bounds, etc.). Fig.~\ref{fig:2dmean2} shows the true structures and those recovered by Bayesian inversion. As expected, we are still able to successfully recover the subducting slab based from the inversions for $L$. Unlike the previous example however, the added complexity brought about by such a type of tectonic setting makes it more difficult for surface wave tomography. One of the reasons for this is the choice of parameterisation as other means of parameterising the model might increase the quality of the recovered images. Other glaring features such as the presence of blobs in $\xi$ instead of sharp edges ascribe to the use of long period observations. Finally, the lateral resolution is very much degraded by random uncorrelated noise being mapped in the parameter space.

\begin{figure}[H]
\centering
\includegraphics[width=0.5\textwidth]{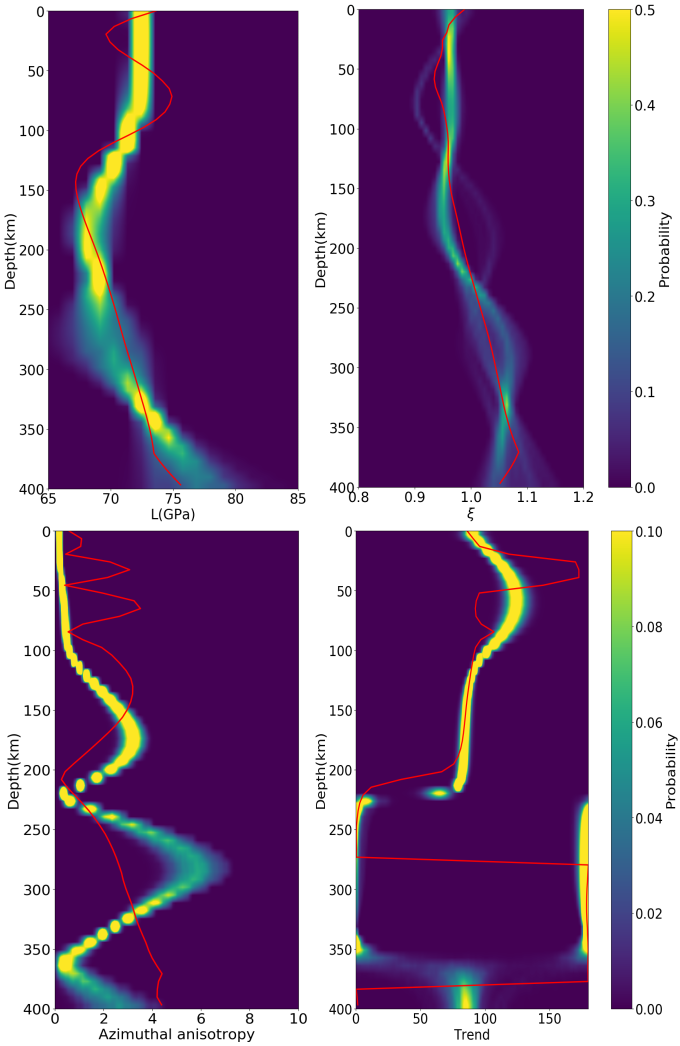}

\caption{\textbf{Subduction case}. 1D marginal posterior distributions of $L$, $\xi$, peak-to-peak azimuthal anisotropy, and its fast azimuth $\Psi$ at a specific geographical location, inferred from the Bayesian inversion of surface wave dispersion curves. The true structures are plotted in solid red.}
\label{fig:1dprofile2}
\end{figure}

Fig.~\ref{fig:1dprofile2} displays the 1D marginal posterior depth profiles of the model parameters at a specific geographical location. As predicted, the sharp gradients such as those shown in $L$ at 75 km are not recovered by surface waves but are spatially-averaged instead. The complex azimuthal anisotropy structure also fails to be rendered by surface wave inversion in part due to the type of parameterisation. It is also quite evident that surface waves lose resolution with depth as initially observed in the sphere case.

\section{Discussion and conclusion}
We have demonstrated conventional anisotropic surface wave tomography. Here, the inverse problem is fully Bayesian in that the solution is an ensemble of models distributed according to a posterior probability. Here, the parameter space is searched using a Markov chain Monte Carlo algorithm with importance sampling. This method was applied to azimuthally-varying surface wave dispersion curves computed from a 3D deforming upper mantle.

\medskip
In both synthetic examples, conventional surface wave tomography was able to recover robust features. However, the resolving power of surface waves limits the vertical resolution of the recovered structures. As exemplified in both cases, surface are long period observations and hence cannot resolve small-scale features. These features are instead spatially-averaged and are smooth as a result. It is also important to emphasize that the choice of parameterisation regularises the inverse problem. As initially stated, the positions of the control points where the values are perturbed are fixed. As a consequence, the results from all the chains inflict strong dependency of the final result on the parameterisation. Such is apparent in the recovered azimuthal anisotropy structures. Finally, since the energy associated with surface waves tend to be more concentrated near the surface, they thus tend to decrease resolution with depth. This is evident in both inversions where below the 250 km mark, the width of the posteriors begin to increase.

\medskip

To handle such complications, possible future avenues include the incorporation of higher-modes. Throughout the inversions, we only considered fundamental-mode surface waves. Using higher modes would increase the sensitivity of surface waves to deeper structures thus providing adequate resolution with depth \citep[e.g.][]{simons1999deep}. Another alternative route we could take is to consider trans-dimensional approaches where the number of model parameters to be inverted for are treated as an unknown. In such cases, the model adapts to the data itself, thus providing a state of balance between model complexity and resolution \citep{bodin2012transdimensional}.

\medskip
Across the horizontal, the recovered structures appear to be less resolved since unlike teleseismic body waves, surface waves exhibit poor lateral resolution. Still, we should expect the horizontal structures to be smooth. This is not the case however as this was a result of using randomly uncorrelated data noise. This random noise maps as small-scale artifacts in the model, which clearly explains the lateral discontinuities. Since real data noise are inherently spatially- and periodically-correlated, it is therefore necessary to build the full data covariance matrix and account for them in the likelihood function.

\medskip

Finally, even though we placed ourselves in the best case scenario, that is, imposing hard \textit{a priori} constraints in the inversions by setting the correct values for $A$, $\phi$, $\eta$, $B_c$, and $B_s$, we are still hindered by the problems associated with conventional surface wave tomography. In practice, these values are treated as unknowns in the inversion or are determined based on empirical relations. Implementing the former complicates the inversion procedure in a way that it increases the model complexity, and that using only one type of data may not resolve these parameters at every location. One is thus forced to go with the latter, where they utilise simple relations such as the $V_pV_s$ ratio to constrain $V_p$-related parameters. Velocity models inferred from such simplistic formulations however may not be representative of the would-be recovered structure especially in situations where complex underlying mechanisms dictate the lithological integrity of the region. The tomographic problem should therefore be approached from a different perspective, where in lieu of prescribing tensor symmetries at the outset it is instead driven by key geophysical processes, such as the work of \citet{10.1093/gji/ggaa577}.

\section*{Acknowledgements}

The provision of the computational resources are made possible through the ERC grant $-$ TRANSCALE: Reconciling Scales in Global Seismology project. The manuscript is part of the PhD dissertation of JK Magali. All computations were carried out through the in-house Transcale cluster situated in Lyon.

\bibliographystyle{plainnat}
\bibliography{bibfile}
\end{multicols}
\end{document}